\documentclass[prx,aps,twocolumn,preprintnumbers,amsmath,amssymb,superscriptaddress,showpacs]{revtex4-1}

\usepackage{amsmath}
\usepackage{graphicx}
\usepackage{amssymb}
\usepackage{color}
\usepackage[bookmarks=true,colorlinks=true,linkcolor=blue,urlcolor=blue,citecolor=blue]{hyperref}
\usepackage{dcolumn}% Align table columns on decimal point
\usepackage{bm}% bold math
\hypersetup{colorlinks=true, pdfstartview=FitV, linkcolor=blue, citecolor=black, plainpages=false, pdfpagelabels=true, urlcolor=blue}
\usepackage[all]{hypcap}
\usepackage{verbatim}

\newcommand{\be}{\begin{equation}}
\newcommand{\ee}{\end{equation}}

\newcommand{\bit}{\begin{itemize}}
\newcommand{\eit}{\end{itemize}}
\newcommand{\bea}{\begin{eqnarray}}
\newcommand{\eea}{\end{eqnarray}}

\usepackage{epsfig}

\begin{document}

\title{Gapless spin liquid in the kagome Heisenberg antiferromagnet with 
Dzyaloshinskii-Moriya interactions}
%\title{Gapless spin-liquid ground state in the $S = 1/2$ kagome Heisenberg 
%antiferromagnet with Dzyaloshinskii-Moriya interactions}
%\title{Effect of Dzyaloshinskii-Moriya Interactions in the Kagome Heisenberg 
%Antiferromagnet}

\author{Chih-Yuan Lee} 
\affiliation{Department of Physics, National Taiwan University, Taipei 10617, 
Taiwan}

\author{B. Normand}
\affiliation{Neutrons and Muons Research Division, Paul Scherrer Institute, 
CH-5232 Villigen PSI, Switzerland}

\author{Ying-Jer Kao}
\affiliation{Department of Physics, National Taiwan University, Taipei 10617, 
Taiwan}
\affiliation{National Center of Theoretical Sciences, National Tsing Hua 
University, Hsinchu 30013, Taiwan}
\affiliation{Department of Physics, Boston University, Boston, MA 02215, USA}
\date{\today}

\begin{abstract}
A preponderance of evidence suggests that the ground state of the 
nearest-neighbor $S = 1/2$ antiferromagnetic Heisenberg model on the 
kagome lattice is a gapless spin liquid. Many candidate materials for the 
realization of this model possess in addition a Dzyaloshinskii-Moriya (DM) 
interaction. We study this system by tensor-network methods and deduce 
that a weak but finite DM interaction is required to destabilize the gapless 
spin-liquid state. The critical magnitude, $D_c/J \simeq 0.012(2)$, lies 
well below the DM strength proposed in the kagome material herbertsmithite, 
indicating a need to reassess the apparent spin-liquid behavior reported 
in this system. 
\end{abstract}

\maketitle

\section{Introduction}
\label{sintro}

Quantum spin liquids have become a central focus of efforts in condensed 
matter physics to understand phenomena including quantum entanglement, 
high-dimensional fractionalization, and topological properties in many-body 
systems \cite{rsb}. In this context, the $S = 1/2$ kagome Heisenberg 
antiferromagnet (KHAF) is one of the most fundamental and controversial 
models to have been studied both theoretically and experimentally 
\cite{qslkagome}. On the theoretical and numerical side, extensive and 
highly refined analyses by a wide range of methods have come out in favor 
of both a gapped \cite{rs,rjws,ryhw,rgfblr,rdms,rjwb,rnsh,rlt,rmchw} and a 
gapless \cite{rrhlw,rhrlw,ribsp,ripb,ripbc,rjkhr,rlxclxhnx,rhzop} spin-liquid 
ground state. However, with the most recent studies by both tensor-network 
\cite{rjkhr,rlxclxhnx} and density-matrix renormalization-group (DMRG) 
\cite{rhzop} methods indicating a gapless U(1) spin liquid, evidence is 
mounting that the physics of the KHAF may be driven by maximizing the 
kinetic energy of gapless Dirac spinons \cite{rlxclxhnx,rrhlw,rhrlw}. 

Experimental approaches to the kagome problem have focused largely on the 
material ZnCu$_3$(OH)$_6$Cl$_2$ (herbertsmithite) \cite{hs}, which offers an 
ideal kagome lattice of $S = 1/2$ Cu$^{2+}$ ions. No evidence has been found 
for long-ranged magnetic order, spin freezing, or indeed a spin gap at 
temperatures as low as 50 mK \cite{rmea,rhea,roea,rhhcnrbl}, which is 3000 
times smaller than the characteristic exchange energy, $J$. However, some 
neutron spectroscopy and nuclear magnetic resonance data have been interpreted 
more recently as showing a small gap \cite{rhnwrhbl,rfihl}. While extensive 
studies have been devoted to the characterization of impurities in 
herbertsmithite \cite{rlkqlhhk,rbnllbdtm,rvkksh,rfea}, it is important 
not to lose sight of the possibility that the physics of the system could 
be influenced by its Dzyaloshinskii-Moriya (DM) interactions. DM terms appear 
naturally in a spin Hamiltonian as a consequence of spin-orbit interactions, 
and cancel only when the bond is a center of inversion symmetry. Thus they 
are ubiquitous in low-symmetry Cu materials, where the square geometry of the 
active $d_{x^2-y^2}$ orbital does not fit the structural geometry, as is the case 
for the triangular motifs of the kagome lattice. An early determination of the 
DM interaction in herbertsmithite by electron spin resonance (ESR) \cite{rzea} 
proposed the values $D_z = 0.08 J$ for the out-of-plane component and $D_\rho
 = 0.01 J$ in-plane, while a subsequent theoretical analysis of the same data 
suggested the bounds $0.044 \le D_z/J \le 0.08$ \cite{rscm}. 

Previous studies of DM interactions in the KHAF were based on exact 
diagonalization (ED) of small clusters. Shortly after the discovery of 
herbertsmithite, efforts were made \cite{rrs1,rrs2} to interpret powder 
susceptibility measurements at intermediate and higher temperatures on the 
basis of linked-cluster expansions and ED calculations using clusters of 12 
and 15 sites. While both in- and out-of-plane DM components were considered, 
few definite conclusions were possible due to uncertainty over the impurity 
contributions to the experimentally measured quantities. ED studies of the 
ground state were performed both by considering the pure system on clusters 
of up to 36 sites \cite{rcfll} and by considering the model in the presence 
of a single impurity on clusters of up to 26 sites \cite{rrmlnm}; partly 
out of numerical convenience, these calculations were performed using 
only an out-of-plane DM term. ED is known to predict a gapped (Z$_2$) 
spin-liquid ground state, although the magnitude of this gap has never been 
agreed upon \cite{rsl} and the result is now suspected to be an artifact 
of the small cluster size \cite{rlsm}. This gapped state is naturally 
robust against small $D_z$, and a transition to the 120$^{\rm o}$ magnetic 
order favored by the out-of-plane DM term was found at $D_c \simeq 0.10 J$ 
\cite{rcfll,rrmlnm}. Schwinger-boson methods, which also favor a gapped 
ground state, have been used \cite{rmcl,rhfs} to obtain similar $D_c$ values, 
subject to uncertainties over representative values of $S$ and $N$ [for SU($N$) 
spins] to employ in this framework. Although the same methods have also been 
used to suggest that $D_z$ may induce a chiral state \cite{rmblb}, DMRG studies 
\cite{rzgs} find only proximity to the chiral spin liquid. This latter 
analysis obtained $D_c = 0.08 J$ while interpreting the ground state at 
$D < D_c$ as a gapless spin liquid. A recent functional renormalization-group 
(FRG) analysis \cite{rhr}, which also was later shown to suggest that the 
ground state of the KHAF is a gapless Z$_2$ spin liquid \cite{rhsir}, has 
again obtained a similar result, $D_c = 0.12(2) J$.

In contrast to these earlier studies, we employ a numerical technique, 
specifically a tensor-network ansatz based on projected entangled simplex 
states (PESS), known to provide a gapless spin-liquid ground state when $D
 = 0$. We will show that, despite its lack of a ``protective'' gap, or even 
a known protective topology, the U(1) spin liquid persists to small but finite 
values of the out-of-plane DM interaction. 

The structure of this manuscript is as follows. In Sec.~\ref{smm} we introduce 
the model and summarize the tensor-network methods we use to analyze it, 
focusing on the technical developments we have introduced in the present 
calculations. In Sec.~\ref{shusimi} we discuss the Husimi lattice, for which 
extremely accurate PESS calculations are possible, to fulfil the dual roles 
of benchmarking the extrapolation of our numerical data and of benchmarking 
the physics of our kagome results. In Sec.~\ref{sem} we present the energy 
and magnetization of the kagome lattice for different values of the DM 
anisotropy. We use the Husimi benchmark in Sec.~\ref{spd} to deduce the 
critical DM coupling strength for the suppression of spin-liquid behavior 
and close in Sec.~\ref{sd} by commenting on the theoretical aspects and 
experimental context of our findings. 

\section{Model and Methods}
\label{smm}

To investigate the effects of DM interactions in the KHAF, we consider the 
model 
\begin{equation} 
H = \sum_{\langle ij \rangle} J {\vec S}_i \! \cdot \! {\vec S}_j + D_z {\hat z}
 \! \cdot \! ({\vec S}_i \times {\vec S}_j).  
\label{eh}
\end{equation} 
For consistency with previous studies \cite{rcfll,rrmlnm,rhr}, and also 
motivated by the ESR analysis \cite{rzea}, we consider only an out-of-plane 
DM component, $D_z$ (henceforth $D$). Following the results of 
Ref.~\cite{rlxclxhnx}, that the gapless spin-liquid regime exists over 
only a narrow range of next-neighbor coupling, we do not consider any 
further-neighbor Heisenberg terms. Following the logic expressed in 
Ref.~\cite{rzea} on the basis of the $g$-factor, that anisotropies in the 
exchange interaction are smaller than DM terms by an order of magnitude 
[in $(g - 2)/g$, which is approximately 0.1 in herbertsmithite], we do not 
consider any XXZ character in Eq.~(\ref{eh}). However, we comment that 
exchange anisotropies have been estimated from ESR data in the candidate 
kagome material vesignieite \cite{rzbotbwm}. 

Numerical methods based on tensor-network representations of the quantum 
many-body wave function \cite{rnkz,rnhomag,rvc} have matured only recently 
to the point at which they can be relied on for the quantitative analysis 
of problems at the leading edge of research in strongly correlated systems 
\cite{rtr1,rtr2}. As generalized matrix-product states, tensor-network wave 
functions obey the area law of entanglement \cite{recp}. Of key importance in 
the kagome problem, their structure allows a real-space renormalization-group 
approach by which one may access the limit of infinite lattice size. In 
overall structure, our calculations follow the approach of Refs.~\cite{rxea} 
and \cite{rlxclxhnx}, summarized in the remainder of this section, but differ 
in a number of details and are performed using code we have written based on 
the Uni10 tensor-network library (https://uni10.gitlab.io/) \cite{uni10}.

\begin{figure}[t]
\centering
\includegraphics[width=0.96\columnwidth]{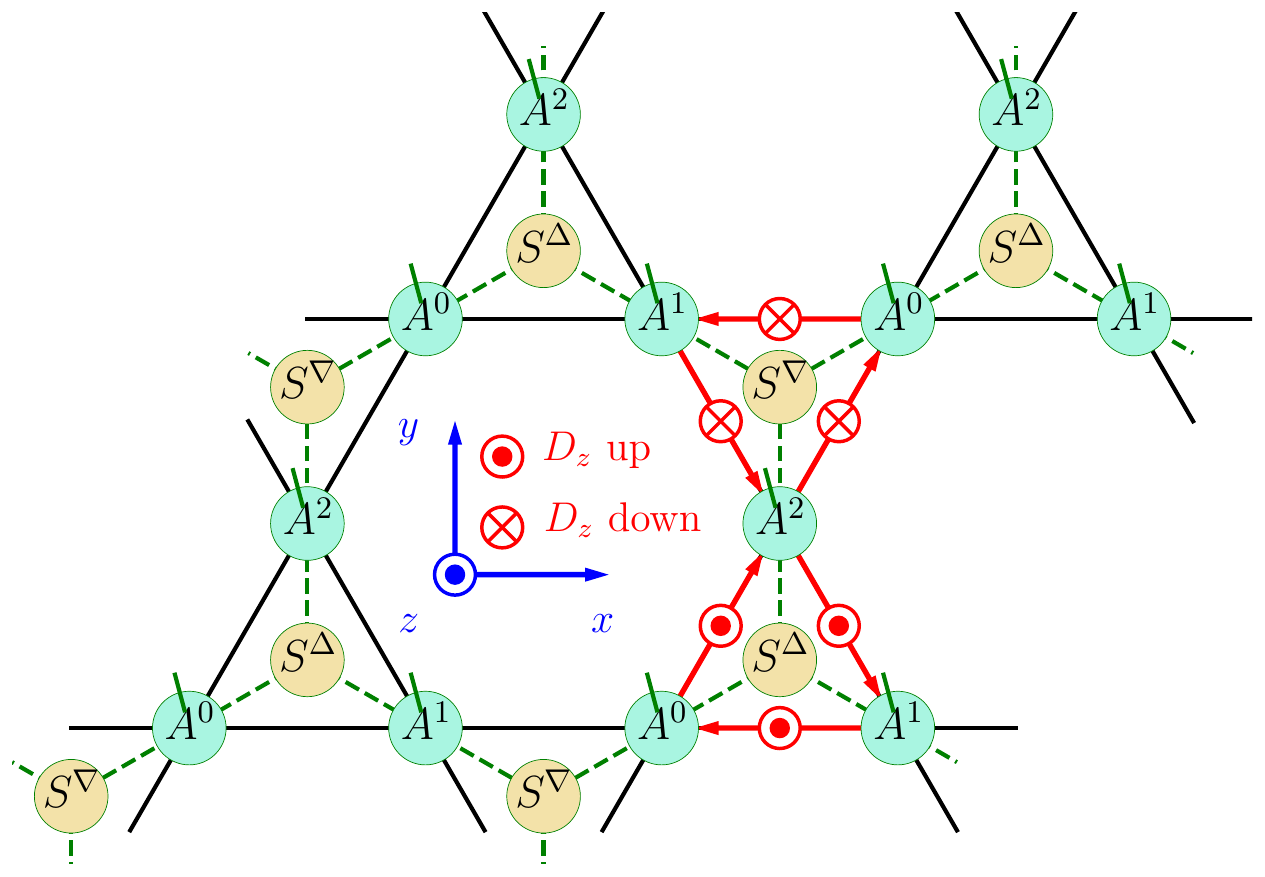}
\caption{Graphical representation of the kagome lattice, the geometry of the 
DM interactions in Eq.~(\ref{eh}), and the 3-PESS ansatz. $S^{\Delta}$ and 
$S^{\nabla}$ are the simplex tensors, which encode the multipartite entanglement 
of the kagome triangles, and $A^{0}$, $A^{1}$, $A^{2}$ are projection tensors.}
\label{fig:PESS ansatz}
\end{figure}

\subsection{PESS as a wave-function ansatz}

The crucial element of PESS \cite{rxea}, which goes beyond the conventional 
pairwise projected entangled pair states (PEPS) construction \cite{rtr2}, is 
its ability to capture the nontrivial multipartite entanglement within each 
lattice unit, or simplex \cite{ra,rxea,rlea}, of a frustrated quantum spin 
system. By the geometry of the kagome lattice, and of the nearest-neighbor 
interactions (denoted $\langle ij \rangle$) in Eq.~(\ref{eh}), the system is 
described naturally by the ``3-PESS'' shown in Fig.~\ref{fig:PESS ansatz}, 
where $S^{\Delta}$ and $S^{\nabla}$ are the two types of three-site simplex tensor 
and $A^{0}$, $A^{1}$, and $A^{2}$ are projection tensors. The dangling solid 
lines in Fig.~\ref{fig:PESS ansatz} denote the physical (spin) degrees of 
freedom, which add a physical tensor dimension $d = 2$ for $S = 1/2$. The 
inter-tensor bonds are virtual objects whose tensor bond dimension, which 
here we denote by $\chi$, sets the maximal number of virtual states that 
can be kept within the ansatz and functions as the truncation parameter in 
the tensor-network representation. We impose translational symmetry and hence 
Fig.~\ref{fig:PESS ansatz} corresponds to the mathematical expression of the 
wave function 
\[ | \Psi \rangle = \text{Tr} (... S^{\alpha}_{a'b'c'} A^{0}_{a'a,\sigma_{i}} 
A^{1}_{b'b,\sigma_{j}} A^{2}_{c'c,\sigma_{k}} ...) |... \sigma_{i} \sigma_{j} 
\sigma_{k} ... \rangle, \] 
where $\alpha = \Delta,\nabla$ denote two simplex tensors for up- and 
down-triangles of the lattice, $\{\sigma_{i},\sigma_{j}, \dots\}$ denote the 
spin basis states on lattice sites $i$, $j$, \dots, and $\{a,b,\dots\}$ 
denote the $\chi$ virtual bond states.

\begin{figure}[t]
\centering
\includegraphics[width=\columnwidth]{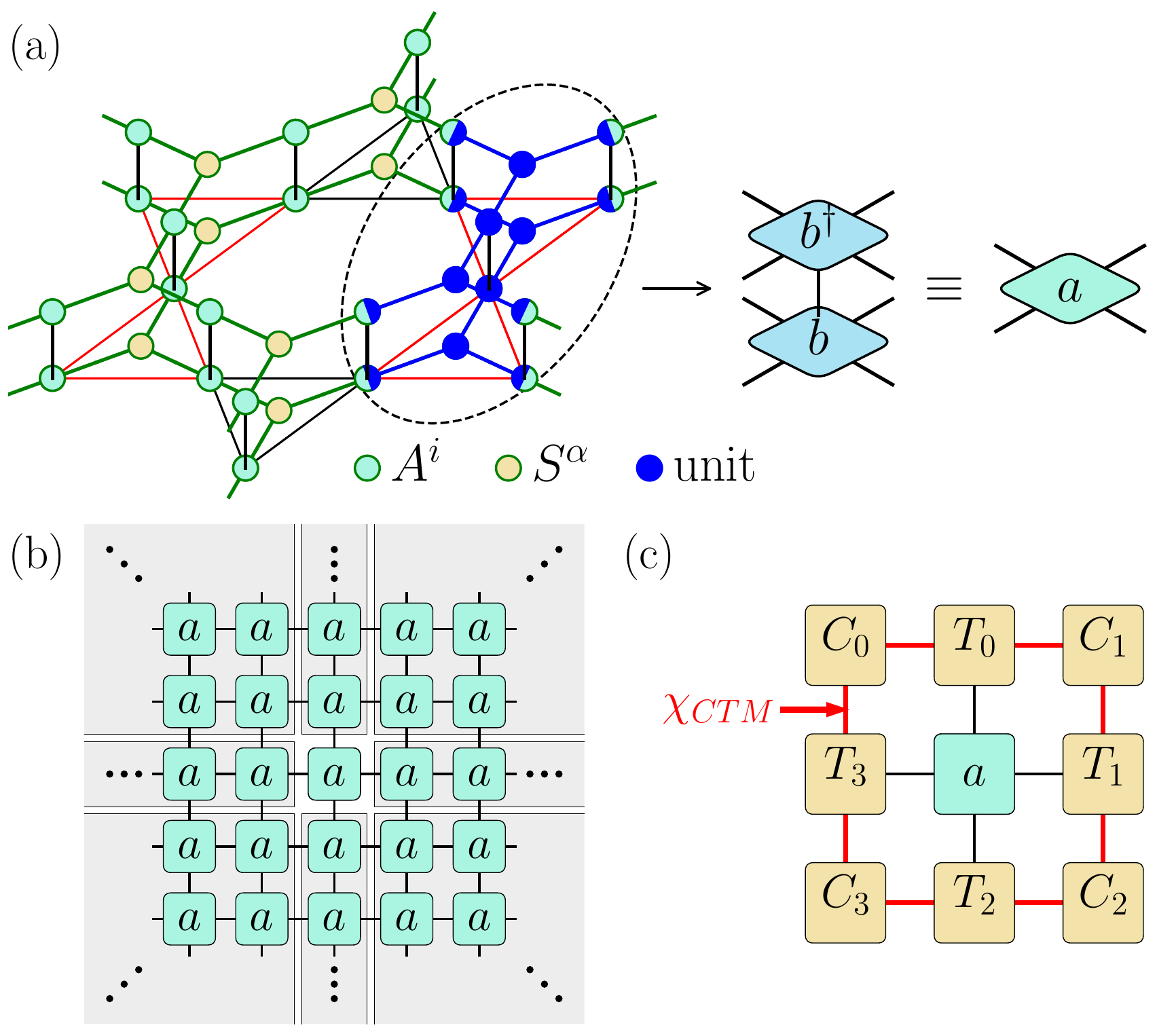} 
\caption{(a) Representation of a bilayer of 2D infinite tensor networks, 
whose repeat units may be combined into a single tensor, $a$, and whose 
contraction is required in the calculation of a physical expectation value. 
(b) Illustration of the square lattice of $a$ tensors and the blocking scheme 
adopted in the CTM approach. (c) Approximation to the environment of a single 
$a$ tensor on the square lattice by four $C$ and four $T$ tensors, each with 
boundary bond dimension $\chi_{CTM}$.} 
\label{fig:ctm}
\end{figure}

A PEPS/PESS representation is manipulated efficiently by an imaginary-time 
projection technique \cite{rjwx} similar to the infinite time-evolving 
block-decimation method \cite{rv,rov}, which we apply to project out the 
ground-state wave function. By decomposing Eq.~(\ref{eh}) into $H = H_{\Delta}
 + H_{\nabla}$, where the two terms contain respectively all Hamiltonian terms 
on up- and down-pointing triangles, we approach the optimized PESS ground 
state by applying $e^{-\tau H_{\Delta} }$ and $e^{-\tau H_{\nabla}}$ successively on a 
random initial wave function, where $\tau$ is a small imaginary time step. 
We comment that this procedure breaks the threefold symmetry of the kagome 
system, with the result that evaluations of the two-triangle unit are 
actually performed on the square lattice, as depicted schematically in 
Fig.~\ref{fig:ctm}(a). We verify the restoration of threefold symmetry in 
the limit of small $\tau$ and large $\chi$ during our calculations of physical 
expectation values \cite{rxea}. 

The projection operators $e^{-\tau H_{\alpha}}$ act on the full triangular simplex 
(schematically $S^{\alpha} A^0 A^1 A^2$) to produce a contracted tensor with 
dimension $(d \chi)^3$ at each step. For the truncation of this tensor, we 
work exclusively at the level of the simple-update method \cite{rjwx,simple_2,
rtr1}, based on local tensor contractions and explained in detail for the 
kagome lattice in Ref.~\cite{rxea}; this approach has been found to yield the 
optimal PESS ground states based on efficiency of convergence and accessible 
$\chi$ values. We comment that the simple-update treatment is essentially 
complete on the Husimi lattice \cite{rlea}, where the simplex tensors have no 
connection other than their local bonds, and we will exploit this property 
in Sec.~\ref{shusimi} to assist in interpreting our kagome calculations. 

\begin{figure}[t]
\centering
\includegraphics[width=\columnwidth]{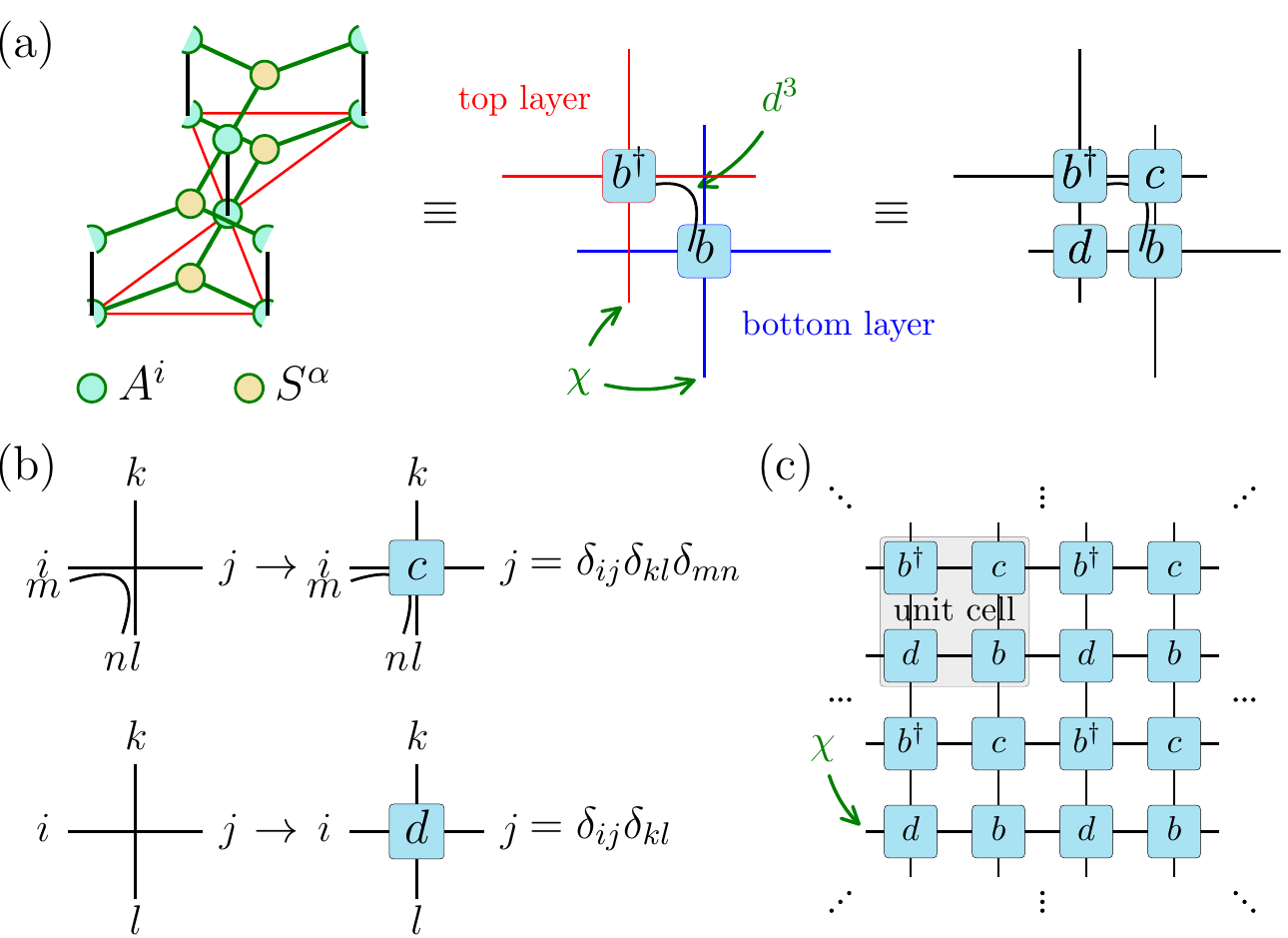}
\caption{Representation of the dimension-reduction procedure. (a) The 
calculation of an expectation value is the contraction of two tensor 
networks whose minimal unit is $b^{\dagger}$ in the top layer and $b$ in 
the bottom layer, the two being connected by contraction of the physical 
bond. To adopt the one-layer CTM method, we introduce additional contraction 
tensors, $c$ and $d$. (b) Pictorial definition of the tensors $c$ and $d$. 
(c) Resulting reduction of the tensor network in Fig.~\ref{fig:ctm}(b), 
which has bond dimension $\chi^{2}$ and one-tensor unit cell, to a network 
with bond dimension $\chi$ and four-tensor unit cell.}
\label{fig:red}
\end{figure}

\subsection{Computing expectation values by CTM}

The PESS wave function we obtain is an infinite two-dimensional (2D) tensor 
network. For the calculation of physical expectation values, $\langle \Psi 
| Q | \Psi \rangle$, it is necessary to contract this network, or more 
specifically its ``square,'' represented in Fig.~\ref{fig:ctm}(a). A number 
of approaches exist for this procedure, specifically the use of boundary 
matrix-product states (bMPS) \cite{rv,rov}, which are used to perform 
successive 1D contractions \cite{rxea,rlxclxhnx}, of corner-transfer-matrix 
(CTM) methods \cite{ctm_Nishino,ctm_Vidal}, which proceed directly in 2D, 
and the hybrid method of channel environments \cite{channel}. Here we have 
adopted the CTM scheme, in which the original problem based on tensors $a$ 
[Fig.~\ref{fig:ctm}(b)] is approximated by $C$ and $T$ tensors as shown in 
Fig.~\ref{fig:ctm}(c), and the accuracy of the approximated environment is 
controlled by the boundary bond dimension, $\chi_{CTM}$. The $C$ and $T$ 
tensors are deduced from $a$ and from isometry operations \cite{isometry_1,
isometry_2,isometry_3} by constructing an iterative renormalization scheme 
based on the invariance of the system under the addition of rows and columns. 
Technically, it is necessary to store the environments for all tensors within 
the unit cell during the iteration. 

To maximize the $\chi$ value for which we can compute physical quantities, we 
follow a recent proposal \cite{rxlhxclx} for optimizing the tensor contraction 
process. This method, originally proposed to optimize bMPS contractions and 
employed in Ref.~\cite{rlxclxhnx} to extend the maximum $\chi$ attainable on 
the kagome lattice from $15$ to $25$, can also be applied within a CTM 
approach. Its essence is to transform the original calculation, which is 
the contraction of a double-layer tensor network with bond dimension 
$\chi^{2}$ and a one-tensor unit cell [Fig.~\ref{fig:red}(a)], to the 
contraction of a single-layer network with a 2$\times$2 unit cell, as 
represented in Fig.~\ref{fig:red}(c). The upper ($b^{\dagger}$) and lower 
($b$) tensor networks are combined by introducing the tensors 
[Fig.~\ref{fig:red}(b)]  
\begin{eqnarray*}
c_{ijklmn} & = & \delta_{ij} \delta_{kl} \delta_{mn}, \\
d_{ijkl} & = & \delta_{ij} \delta_{kl}.
\end{eqnarray*}

\begin{figure}[t]
\centering
\includegraphics[width=\columnwidth]{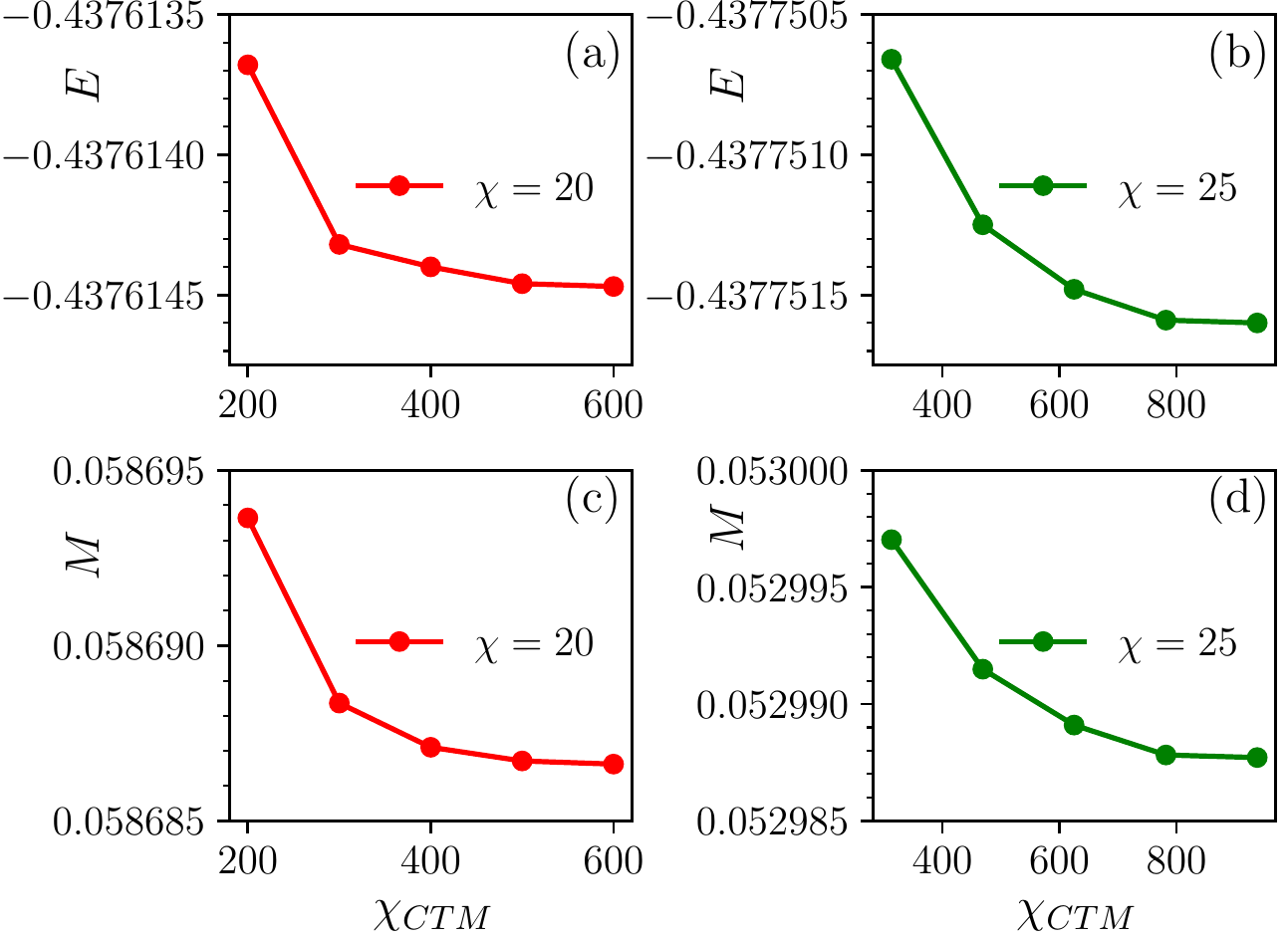}
\caption{Convergence of the ground-state energy, $E$ (a,c), and staggered 
magnetization, $M$ (b,d), as functions of the boundary bond dimension, 
$\chi_{CTM}$, for bond dimensions $\chi = 20$ (a,b) and 25 (c,d).}
\label{fig:boundary_dim}
\end{figure}

Because CTM is an approximate contraction method, in which the error is 
controlled by $\chi_{CTM}$, the variational principle is not applicable and 
any physical expectation value may increase or decrease with increasing 
$\chi_{CTM}$. The expectation values on which we focus here are the 
ground-state energy per site, $E = {\textstyle \frac{1}{6}} (E_\Delta + 
E_\nabla)$, and the staggered magnetization (ordered moment per site), 
\begin{equation}
M = {\textstyle \frac{1}{3}} \sum_{i = 1,2,3} \sqrt{ \langle S_i^x \rangle^2
 + \langle S_i^y \rangle^2 + \langle S_i^z \rangle^2}, 
\label{esm}
\end{equation}
where $i$ denotes the three sites of an up- or down-simplex in the 
translationally invariant infinite system, and as noted above all three 
sites and the two simplex types become equivalent in the limits of small 
$\tau$ and large $\chi$ and $\chi_{CTM}$. To analyze convergence as a function 
of $\chi_{CTM}$, in Fig.~\ref{fig:boundary_dim} we show the evolution of $E$ and 
$M$ of the KHAF for a representative DM interaction $D = 0.016$ and for $\chi$ 
values of 20 and 25. We observe that both $E$ and $M$ show well-controlled 
convergence with increasing $\chi_{CTM}$ and that, beyond the value $\chi_{CTM} 
\approx \chi^2$, where the errors due to the finite value of $\chi_{CTM}$ are 
expected to be small compared with those due to the finite $\chi$, the 
relative changes in both quantities are extremely small. 

\section{Husimi HAF with DM Interactions}
\label{shusimi}

The nature of our PESS results is that we obtain physical expectation values, 
specifically $E$ and $M$ of the preceding section, for a series of finite 
values of the truncation parameter, which is the tensor bond dimension, $\chi$. 
The physics of the true ground state at any given value of $D$ is obtained by 
extrapolating this series to the limit of infinite $\chi$. To illustrate the 
nature of this extrapolation in the most systematic way possible, we turn to 
the Heisenberg antiferromagnet, with DM interactions, on the Husimi lattice. 

\begin{figure}[t]
\centering
\includegraphics[width=0.96\columnwidth]{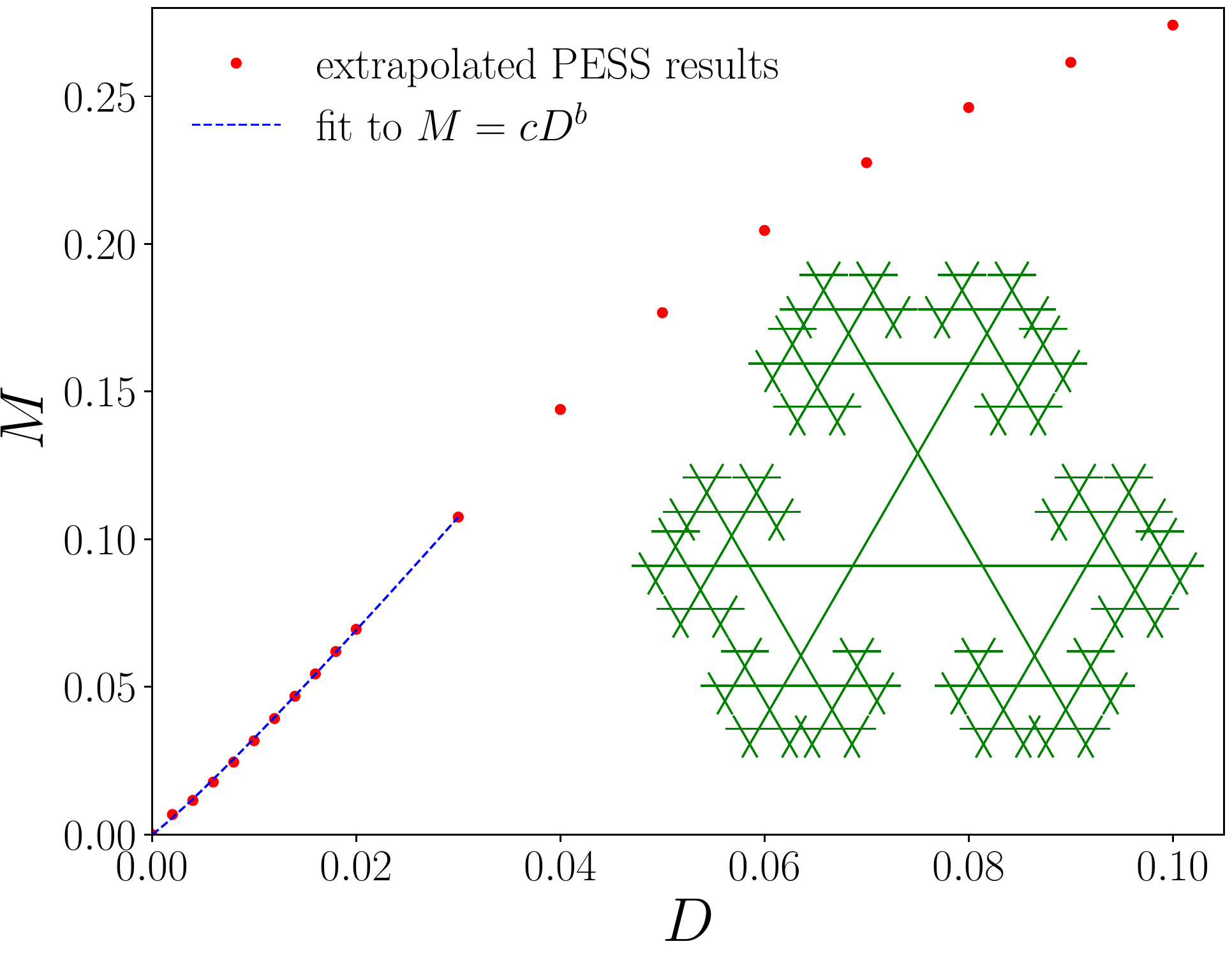}
\caption{$M$ as a function of $D$ on the Husimi lattice, showing a near-linear 
scaling with proportionality constant $c = 5.2(2)$ and exponent $b = 1.10(2)$. 
Inset: Husimi lattice.}
\label{fig:mdh}
\end{figure}

The Husimi lattice, whose geometry we show in the inset of Fig.~\ref{fig:mdh}, 
is a Bethe lattice of corner-sharing triangles. It possesses the same local 
coupled-triangle physics as the kagome lattice, but all longer paths that 
connect spins in the KHAF are entirely absent. We consider only the infinite 
Husimi lattice, which possesses translational invariance and is distinct from 
the ``Husimi tree,'' a finite system in which half of the sites are located 
on the boundary, leading to some singular properties \cite{rlea}. 

Physically one may anticipate that the Husimi HAF is a ``less frustrated'' 
quantum spin model than the KHAF, giving a lower tendency to spin-liquid 
formation. From the viewpoint of PESS calculations, the absence of all longer 
loops means that the simple-update method is effectively complete for the 
Husimi lattice \cite{rlea}, in the sense that no additional measures are 
required to account for longer paths, for example in the calculation of 
bond environment terms \cite{rxea}. As a result, the problem remains at the 
level of a local minimization, making it possible to reach very large values 
of $\chi$ and hence to perform very reliable extrapolation of all physical 
expectation values to the large-$\chi$ limit. 

\begin{figure}[t]
\centering
\includegraphics[width=1.0\columnwidth]{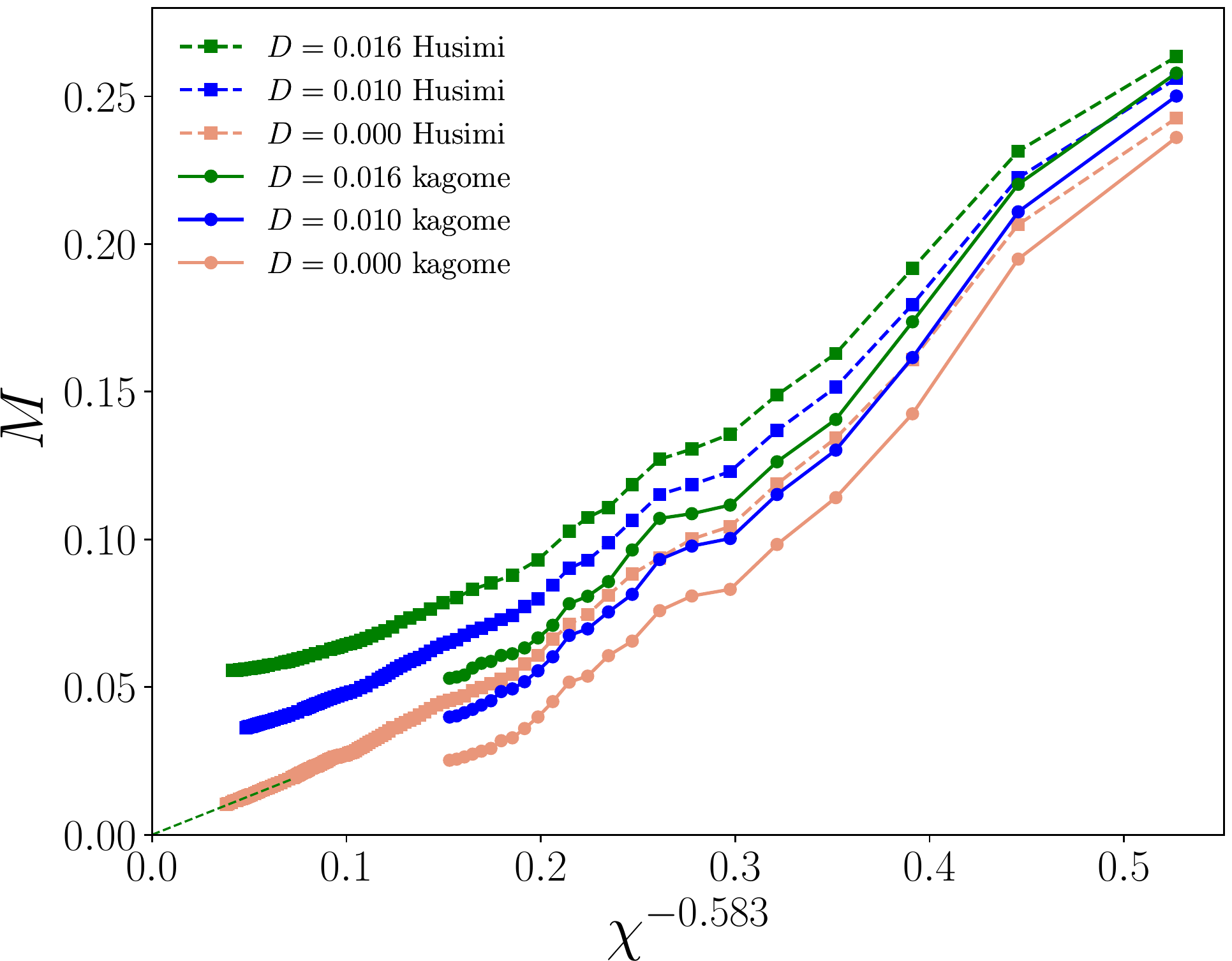}
\caption{Extrapolated $M$ as a function of $\chi$ for the Husimi and kagome 
systems at three different values of the DM interaction.}
\label{fig:hkem}
\end{figure}

We focus for illustration on $M(\chi)$ in the Husimi HAF, which we show in 
Fig.~\ref{fig:hkem} for three different values of $D$. We stress that $M$ 
is always finite in PESS calculations on the Husimi and kagome lattices at 
finite values of $\chi$ \cite{rlxclxhnx}, and that only reliable extrapolation 
to infinite $\chi$ can be used to determine whether or not the magnetic order 
is real; however, a real 120$^\circ$-ordered antiferromagnetic phase is expected 
on both lattices at larger values of $D$ \cite{rcfll,rrmlnm}. Guided by the 
possibility of gapless spin-liquid or antiferromagnetic phases at infinite 
$\chi$, we consider only power-law fitting forms. With many data points 
available up to $\chi = 280$, we obtain reliable fits and accurate intercepts. 
When $D = 0$, we obtain an accurate fit to $M \propto \chi^{-a}$, i.e.~we find 
in accord with Ref.~\cite{rlea} that $M = 0$ in the limit of infinite $\chi$ 
and that the exponent is $a = 0.583(5)$. For all finite values of $D$, we 
find that $M(\chi \rightarrow \infty)$ is finite, meaning that the gapless 
spin-liquid phase is present in the Husimi HAF only at $D = 0$.

To consider the evolution of the physical properties of the system with $D$, 
and with a view to examining the nature of a possible quantum phase transition 
on the kagome lattice from an ordered antiferromagnetic phase at higher $D$ to 
a quantum spin liquid \cite{rcfll,rrmlnm}, we illustrate the situation for the 
Husimi lattice. Because it is not clear that data points at high $D$ should 
fall in the quantum critical regime, we adopt a windowing procedure where we 
fit different numbers of data points (starting at point $(0,0)$, to which we 
ascribe zero error bar \cite{rlea}) and take the fit with the lowest reduced 
chi-squared value as our best estimate. As shown in Fig.~\ref{fig:mdh}, 
we deduce that $M = c D^b$ with $b = 1.10(2)$, implying a nearly, but not 
exactly, linear relation between the ordered moment and the DM interaction 
away from the critical (gapless) phase at $D = 0$.  

In summary, the Husimi HAF provides crucial qualitative and quantitative 
benchmarks for our PESS calculations, showing that magnetically ordered states 
have the lowest energies for spatially infinite systems at finite $\chi$, and 
that $E(\chi)$ and $M(\chi)$ are algebraic. The vanishing of $M(\infty)$ at 
$D = 0$ presents a reliable example of a gapless spin liquid, and the finite 
$M(\infty)$ at all finite $D$ may be expected from the somewhat pathological 
Bethe-lattice geometry \cite{rlxclxhnx}. We caution that the functional 
forms we deduce from our Husimi data do not apply at small $\chi$ 
(Fig.~\ref{fig:hkem}) and that by this measure all of our kagome results 
are in the small-$\chi$ regime, which will determine our treatment of the 
kagome data in Sec.~\ref{spd}. 

\section{KHAF Energy and Magnetization}
\label{sem}

Turning now to the KHAF, it was shown in Ref.~\cite{rlxclxhnx} that 
simple-update PESS calculations up to $\chi = 25$ yield a ground-state 
energy for the nearest-neighbor model, $E(\chi \rightarrow \infty)$, that 
lies below the values obtained from all other techniques (apart from DMRG 
calculations for certain cylindrical geometries). $E(\chi)$ was found to obey 
an algebraic convergence with $\chi$, as on the Husimi lattice \cite{rlea}, 
indicating a gapless ground state \cite{rpvvt}. Also as on the Husimi lattice, 
the PESS wave function was found to have a finite 120$^\circ$ magnetic order 
at all finite $\chi$ values, but with the algebraic $M(\chi)$ lying well 
below the analogous Husimi value. 

\begin{figure}[t]
\begin{center}
\includegraphics[width=0.96\columnwidth]{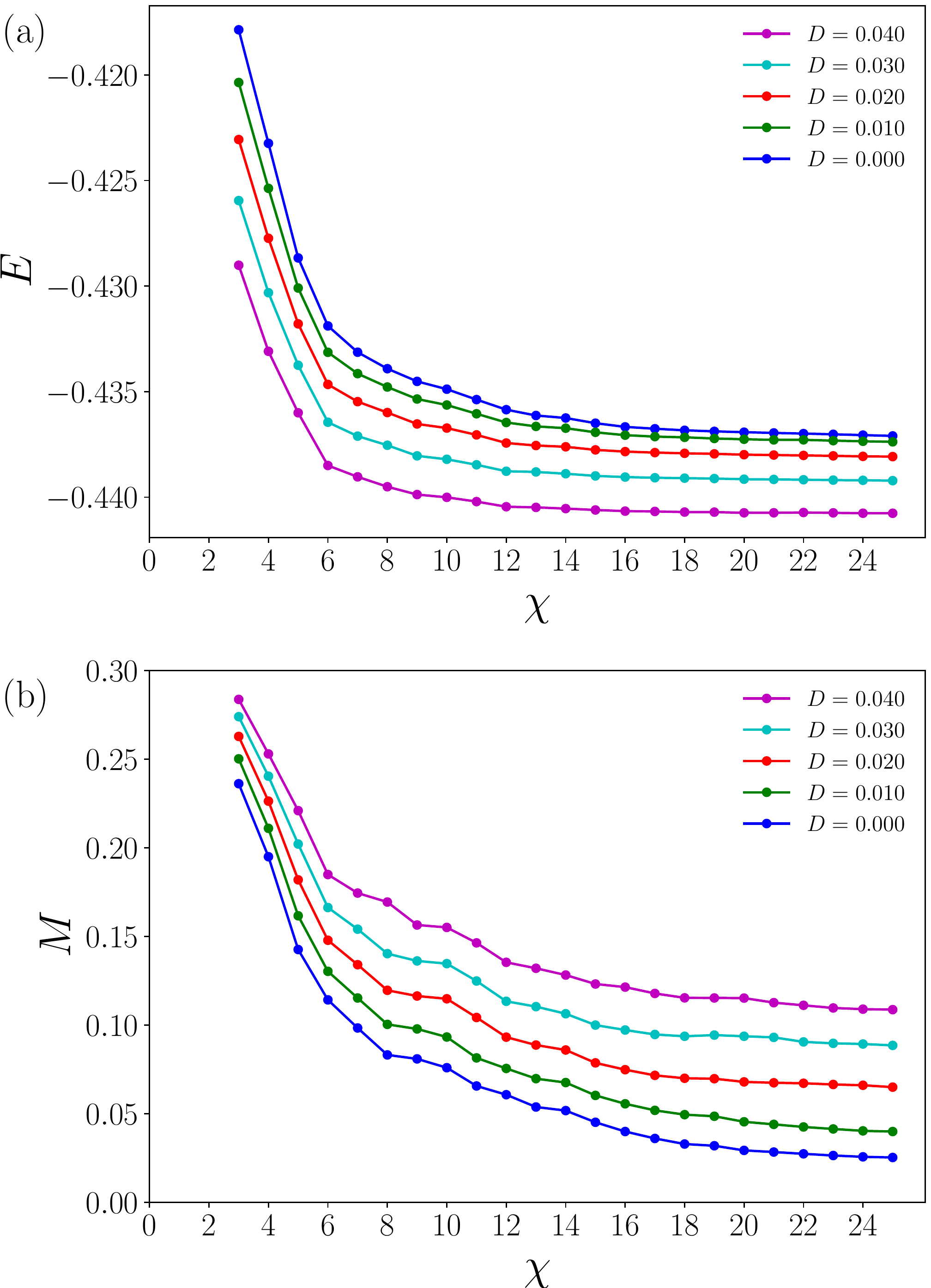}
\end{center}
\caption{KHAF with DM interactions. (a) $E$ as a function of $\chi$, shown 
for three different values of $D$. (b) $M$ as a function of $\chi$, shown 
for three different values of $D$.}
\label{fkdm}
\end{figure}

In Fig.~\ref{fkdm} we extend these results to include DM interactions. 
It is clear that finite $D$ values push down the ground-state energy in 
a monotonic manner [Fig.~\ref{fkdm}(a)], implying a relief of the kagome 
frustration. Equally clear is that $M(\chi)$ is pushed upwards by the effect 
of $D$, in the same monotonic manner, implying that that trend is towards a 
magnetic state, as already found in the Husimi case. Although the 120$^\circ$ 
state on each triangle remains frustrated both for the Heisenberg term and 
for the DM term, it appears that this frustration is lower than that inherent 
in the gapless spin liquid. However, our kagome results terminate at $\chi 
 = 25$, a bond dimension reachable only at great computational cost. As 
implied in Sec.~\ref{shusimi}, and made clear by Fig.~\ref{fig:hkem}, the 
data continue to show artifacts at these $\chi$ values that prevent a 
reliable extrapolation. 

\section{Phase Diagram}
\label{spd}

To make progress under these circumstances in understanding the physics 
of the KHAF, we continue to exploit the comparison with the Husimi system. 
In Ref.~\cite{rlxclxhnx} the stability of the gapless spin-liquid phase was 
investigated by adding a next-neighbor coupling, $J_2$, to the KHAF, and 
stability was demonstrated over a finite, if narrow, regime of $J_2$ by 
comparing both $E$ and $M$. Here we adopt an analogous procedure to investigate 
the effect of $D$ on the gapless spin liquid by comparing our magnetization 
results. 

\begin{figure}[t]
\begin{center}
\includegraphics[width=1.0\columnwidth]{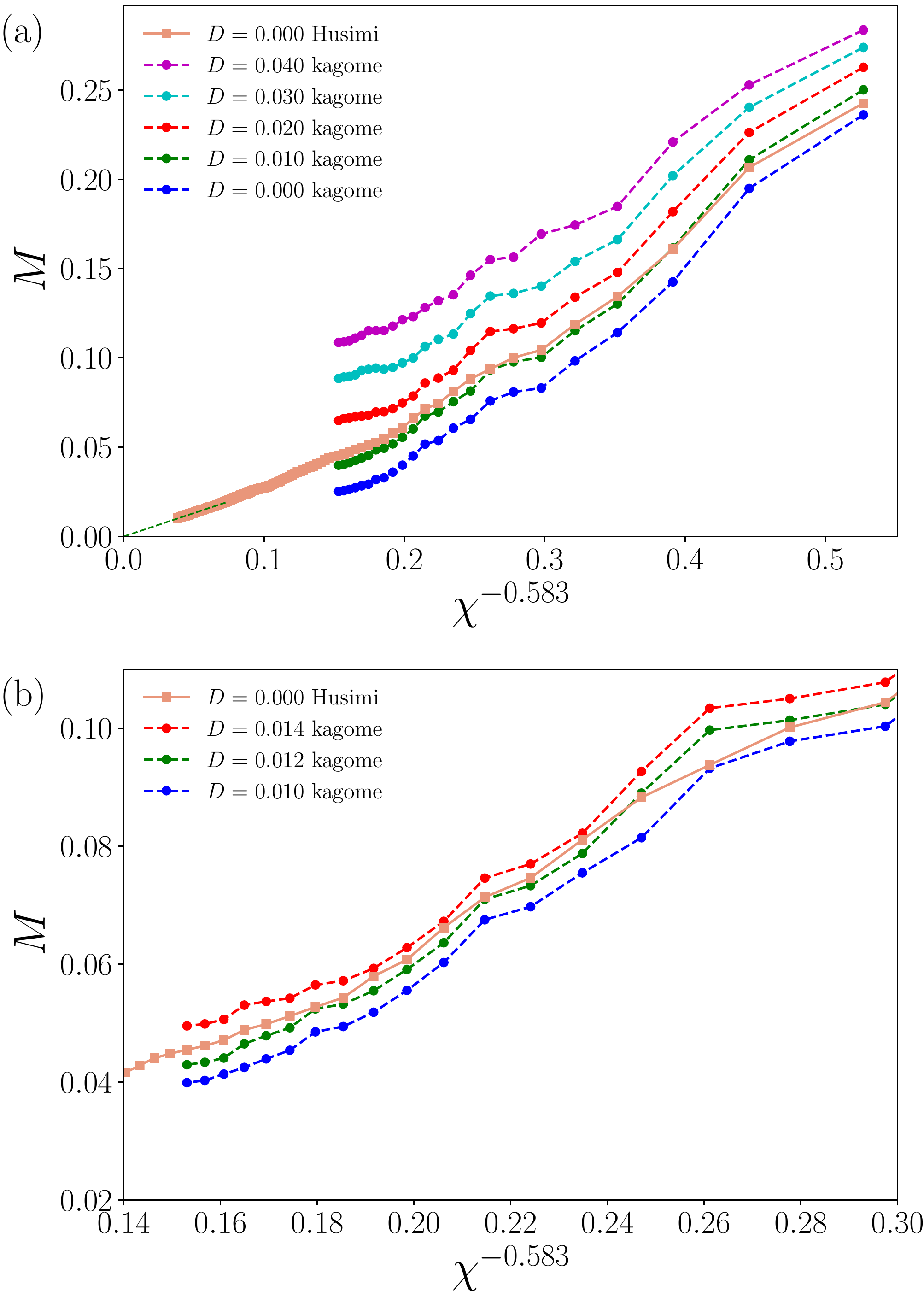}
\end{center}
\caption{Staggered magnetization of the KHAF compared with the Husimi HAF. (a) 
$M(\chi)$, shown as a function of $1/\chi^{0.583}$, calculated for the Husimi 
lattice with $D = 0$ and for the kagome lattice with several values of $D$. 
(b) Detail of $M(\chi)$ allowing the upper limit of the gapless spin-liquid 
phase to be established.} 
\label{fmhk}
\end{figure}

In Fig.~\ref{fmhk}(a) we compare the magnetization of the KHAF, $M(\chi)$, 
computed for several values of $D$, with $M(\chi)$ for the Husimi HAF at 
$D = 0$. Because the Husimi system marks the upper limit of spin-liquid 
behavior, we assert that all magnetization curves lying below this one 
correspond to the parameter regime in which the KHAF has a spin-liquid 
ground state. By its algebraic nature, this spin liquid will be gapless 
for all $D$ values below the critical one where order sets in. The monotonic 
rise of $M(\chi)$ with $D$ in the KHAF [Fig.~\ref{fkdm}(b)] ensures that there 
is only one crossing point, $D_c$, beyond which the finite-$D$ kagome result 
lies above the $D = 0$ Husimi one. By establishing the smallest interval in 
which the $D = 0$ Husimi curve lies between two kagome curves, as shown in 
Fig.~\ref{fmhk}(b), we estimate the error bar on this crossing point and 
hence conclude that $D_c = 0.012(2) J$. We stress that both this value and 
its error are valid within the confines of the comparison of the kagome to 
the Husimi lattice, for which no rigorous theoretical justification exists.

\begin{figure}[t]
\centering
\includegraphics[width=\columnwidth]{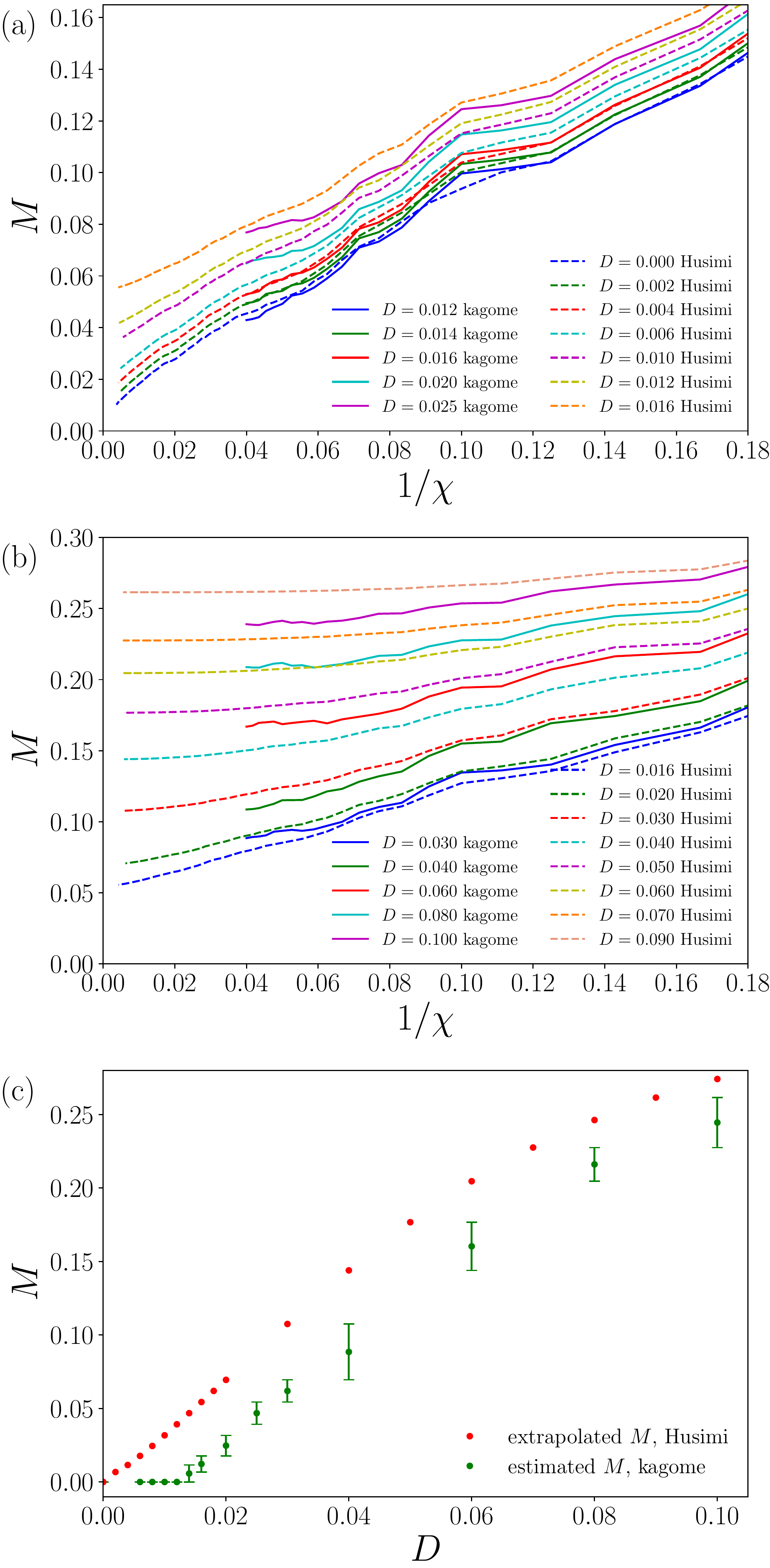}
\caption{Staggered magnetization of the KHAF as a function of $D$. 
(a) $M(\chi)$ for small $D$ values, where the Husimi results can be used to 
extract $M(D)$. (b) $M(\chi)$ for larger $D$ values, where the Husimi results 
no longer provide an accurate constraint. (c) $M(D)$ for the kagome lattice 
compared with the Husimi HAF result of Fig.~\ref{fig:mdh}.}
\label{fmd}
\end{figure}

The key qualitative conclusion of our study is that the gapless spin-liquid 
phase is stable against a finite out-of-plane DM interaction. The persistence 
of the spin-liquid regime, despite the apparent tendency of the DM term to 
drive a 120$^\circ$ antiferromagnetically ordered phase, is evidence both for a 
real physical mechanism underlying the stability of the gapless state and for 
a certain degree of ``protection'' against perturbations. We comment in more 
detail on this issue in Sec.~\ref{sd}. Numerically, the value of $D_c$ we 
find is small compared with values in the literature. While one may worry 
that TNS methods could underestimate this phase boundary by favoring 
magnetically ordered states in the kagome problem, we caution that our 
calculation is really not the same problem as that addressed by ED 
\cite{rcfll,rrmlnm} and Schwinger-boson methods \cite{rmcl,rhfs}, because 
the ground state of the $D = 0$ spin liquid is quite different. By contrast, 
the FRG results of Ref.~\cite{rhsir} for the KHAF without DM interactions are 
similar or identical to ours, and in this case some uncertainty may reside in 
the method used to estimate $D_c$ within the same formalism \cite{rhr}. A 
comparison with variational Monte Carlo results would be helpful in this 
regard. We reiterate the essential point, also made in Ref.~\cite{rzgs} that 
our value of $D_c$ is finite, even when the system has a gapless ground state 
at $D = 0$. 

We conclude the analysis of our data by extracting $M$ as a function of $D$ 
for the KHAF. Using again the Husimi system as a benchmark, we attempt to 
find two $M(\chi)$ curves for the Husimi HAF at different $D$ values which 
completely bracket one $M(\chi)$ curve for the KHAF at fixed $D$. We then use 
the extrapolated $M$ values of the two Husimi results as the upper and lower 
bounds for $M(D)$ on the kagome lattice. This process works rather well at 
small $D$, where the shapes of the $M(\chi)$ curves are quite similar 
[Fig.~\ref{fmd}(a)], but deteriorates at larger $D$, where $M(\chi)$ for the 
KHAF dips more strongly at large $\chi$ [Fig.~\ref{fmd}(b)]. Our best estimate 
for $M(D)$ is shown in Fig.~\ref{fmd}(c), where the growing error bars at 
higher $D$ reflect the kagome-Husimi mismatch. It is clear nevertheless that 
$M$ on the KHAF remains below the Husimi result, tracking it approximately 
after the initial offset. Thus the frustrating effects of the additional, 
longer paths on the kagome lattice appear to be constant, and $D_z$ does not 
act, over the range of our study, to create a regime controlled only by the 
physics of the triangular motifs. 

\section{Discussion}
\label{sd}

Concerning the physics of the gapless spin liquid, the U(1) Dirac-fermion 
state is the first \cite{rrhlw} and still the leading \cite{ribsp,ripb,rjkhr,
rlxclxhnx,rhzop} candidate gapless spin-liquid wave function. Simply from 
the observation that this state is the true ground state, it is tempting to 
suggest \cite{rlxclxhnx} that the mechanism for its stability is to maximize 
the kinetic energy of mobile spinons. An alternative scenario for the 
mechanism is to consider maximizing the contributions from gauge fluctuations 
\cite{rhfb}. However, it has also been suggested, on the basis of ED studies 
of clusters up to 36 sites \cite{rckkcf}, that the kagome point in the phase 
diagram with finite $J_2$ is a change of phase between two different types of 
spin liquid, which could explain the appearance of an anomalously low energy 
scale in the physics of the KHAF. It remains unclear whether such a scenario 
would indicate a real transition between Z$_2$ phases or the presence of a 
U(1) parent phase \cite{rjkhr}. The U(1) Dirac state is known to have 
long-ranged entanglement and a power-law decay of all correlation functions 
\cite{rhrlw}. 

Our results demonstrate that the U(1) state is not immediately disrupted by 
out-of-plane DM interactions, i.e.~these are not strongly relevant from the 
point of view of destabilizing the gapless spin liquid. To the extent that 
the AF-ordered phase is a state of confined spinons, one may conclude that 
the $D_z$ term is not immediately confining. We caution that field-theoretical 
arguments advanced in Ref.~\cite{rhrlw} do suggest that $D_z$ should have an 
immediate effect on the U(1) state, and hence a closer analysis of marginally 
relevant terms is warranted. Because the U(1) Dirac state has no 
well-characterized topology, it is not clear that it could enjoy any 
type of topological protection. It is true that the $D_z$ term does not break 
the U(1) symmetry of the system, which may provide some symmetry protection. 
This is to be contrasted with an in-plane DM interaction, which does break 
the U(1) symmetry and could be expected to promote a finite spin scalar 
product, ${\vec S_i} \! \cdot \! ({\vec S_j} \times {\vec S_k})$, on each 
triangle and hence to favor chiral spin-liquid states. Otherwise the U(1) 
state may enjoy only ``energetic protection'' due to the energy gain of 
its mobile spinons. 

From an experimental standpoint, our results imply that herbertsmithite, 
still by far the best-characterized candidate kagome material, should be 
in a 120$^\circ$-ordered state for the proposed value $D_z/J \approx 0.08$ 
\cite{rzea}. One possible explanation may be that the ESR result, which 
by its nature is an upper bound, is an overestimate. Another, noted in 
Sec.~\ref{spd}, is that our TNS method may be biased towards ordered states, 
and hence the true $D_c$ is larger than our estimate, but it is unlikely that 
our result would contain a factor-10 error. Thus a strong possibility remains 
that herbertsmithite is, after all, magnetically disordered as a consequence 
of its structural disorder, i.e.~this is an extrinsic effect and the system 
is not an intrinsic quantum spin liquid. Discounting this possibility would 
seem to require at minimum a deeper understanding of effects arising due to 
out-of-plane impurities, specifically as regards in-plane polarization 
and interplane coupling.

In summary, we have used tensor-network calculations by the method of projected 
entangled simplex states to demonstrate that the $S = 1/2$ kagome Heisenberg 
antiferromagnet retains a gapless quantum spin-liquid ground state for small 
but finite values of an out-of-plane Dzyaloshinskii-Moriya interaction. Our 
results imply that herbertsmithite should, based on current estimates, lie well 
within an antiferromagnetically ordered phase and thus call for a reassessment 
of the experimental data for this material.

\section*{Acknowledgments}

We thank H.-J. Liao, P. Mendels, J. Reuther, T. Xiang, Z.-Y. Xie, and S. 
Zvyagin for helpful discussions. This work was supported by Ministry of 
Science and Technology (MOST) of Taiwan under Grants No.~105-2112-M-002-023-MY3 
and 104-2112-M-002-022-MY3, and was funded in part by a QuantEmX grant from 
ICAM and by the Gordon and Betty Moore Foundation through Grant GBMF5305 to 
Ying-Jer Kao. We are grateful to the Taiwanese National Center for 
High-Performance Computing for computer time and facilities.

\end{document}